\documentclass[twocolumn,english,aps,prb,showpacs,superscriptaddress]{revtex4-1}
\usepackage[T1]{fontenc}
\usepackage[latin9]{inputenc}
\usepackage{amsmath}
\usepackage{amssymb}
\usepackage{graphicx}
\usepackage{esint}
\usepackage{color}

\makeatletter
\@ifundefined{textcolor}{}
{%
 \definecolor{BLACK}{gray}{0}
 \definecolor{WHITE}{gray}{1}
 \definecolor{RED}{rgb}{1,0,0}
 \definecolor{GREEN}{rgb}{0,1,0}
 \definecolor{BLUE}{rgb}{0,0,1}
 \definecolor{CYAN}{cmyk}{1,0,0,0}
 \definecolor{MAGENTA}{cmyk}{0,1,0,0}
 \definecolor{YELLOW}{cmyk}{0,0,1,0}
}

\makeatother

\usepackage{babel}
\begin{document}

\title{Mechanism of hole propagation in the orbital compass models}

\author{Wojciech Brzezicki }
\affiliation{Marian Smoluchowski Institute of Physics, Jagellonian University,
Reymonta 4, PL-30059 Krak\'ow, Poland }
\affiliation{Institut f\"ur Theoretische Festk\"orperphysik, IFW Dresden, 
D-01171 Dresden, Germany}

\author{Maria Daghofer}
\affiliation{Institut f\"ur Theoretische Festk\"orperphysik, IFW Dresden, 
D-01171 Dresden, Germany}

\author{Andrzej M. Ole\'{s} }
\affiliation{Marian Smoluchowski Institute of Physics, Jagellonian University,
Reymonta 4, PL-30059 Krak\'ow, Poland }
\affiliation{ Max-Planck-Institut f\"ur Festk\"orperforschung, Heisenbergstrasse 1,
D-70569 Stuttgart, Germany }

\date{\today }
\begin{abstract}
We explore the propagation of a single hole in the quantum compass
model, whose nematic ground state is given by mutually decoupled
antiferromagnetic chains. The compass model can be seen as the
strong-coupling limit of a spinless two-band Hubbard model, which we
study here using mean field theory and the variational cluster
approach. Due to the symmetries of the compass model, the inherent
disorder along one lattice direction turns out not to affect hole
motion and doping a hole consequently does not lift the subextensive
degeneracy of the nematic phase.
In order to broaden and deepen understanding, we derive a generalized
itinerant model and address the transition to two-dimensional Ising
order. We observe coherent hole motion in both the nematic and the
antiferromagnetic phases, also in the presence of quantum fluctuations
away from pure Ising exchange. In addition to quantum fluctuations
and interorbital hopping, three-site hopping is found to play an
important role and to dominate propagation in the two-dimensional
Ising limit as well as along the antiferromagnetic chains in the 
nematic order which forms in the compass model.
\end{abstract}

\maketitle

\section{Introduction}


Spin-orbital physics\cite{Nag00,Ole05,Kha05,Ole12} is a very exciting 
and challenging field of research within the theory of strongly 
correlated electrons. Well known examples of Mott insulators with active 
orbital degrees of freedom are two-dimensional (2D) and 
three-dimensional (3D) cuprates,
\cite{Fei97,*Fei98,*Kha97,Brz12,*Brz11,*Brz13} colossal 
magnetoresistance manganites,\cite{Fei99,*Fei05} and vanadates.
\cite{Kha01,*Hor03,*Hor08,*Hor11,*Ave13}
These realistic models are rather complicated and difficult to 
investigate due to spin-orbital entanglement on exchange bonds.
\cite{Ole12,Ole06,You12} A common feature of spin-orbital models is 
intrinsic frustration of the orbital superexchange which follows from 
the directional nature of orbital states and their interactions.
The orbital interactions are frequently considered alone, leading
to orbital ordered states, \cite{vdB99,vdB04,Ryn10} to valence bond
crystal or to orbital pinball liquid exotic quantum states.\cite{Ral12}

We concentrate below first on probably the simplest model that 
describes orbital-like superexchange, the so-called orbital compass
model (OCM),\cite{vdB13} introduced long ago by Kugel and Khomskii.
\cite{Kug82} This 2D model attempts to capture orbital anisotropies
via couplings that are Ising-like along each bond, but where different 
spin components are active along different bond directions. 
A frequently used convention is that interactions take the form 
$J_{x}\sigma_{i}^{x}\sigma_{j}^{x}$ and $J_{z}\sigma_{i}^{z}\sigma_j^z$ 
along the $a$ and $b$ axis of the square lattice. Despite its deceptive 
simplicity, the compass model is challenging even for
classical interactions.\cite{Nus04} Recent interest in this model is 
motivated by its interdisciplinary character as it plays a role in the 
variety of phenomena beyond the correlated oxides; is is also dual to 
recently studied models of $p+ip$ superconducting arrays,\cite{Nus05} 
namely to the Hamiltonian introduced by Xu and Moore,\cite{Xu04}
and to the toric code model in a transverse field.\cite{Vid09} Its
2D and 3D version was studied in the general framework of unified
approach to classical and quantum dualities \cite{Cob10} and in the
2D case it was proven to be self-dual.\cite{Xu04} The OCM was also
suggested as an effective description for Josephson arrays of protected
qubits,\cite{Dou05} as realized in recent experiment.\cite{Gla09}
It could also describe polar molecules in optical lattices and
systems of trapped ions.\cite{Mil07} Recent developments on arrays of 
nitrogen-vacancy centers, constituting point-like defects in a diamond 
matrix,\cite{Gae06,*Neu10} bring a further motivation to the study of 
OCM, as shown in Ref. \onlinecite{Tro12}.

For further discussion of the properties of the 2D OCM it is helpful
to recall the one-dimensional (1D) case. The 1D generalized variant of 
the compass model with $z$-th and $x$-th spin component interactions 
that alternate on even/odd exchange bonds is strongly frustrated, 
similar to the 2D OCM. 
The 1D OCM can be solved exactly by analytical methods in two 
different ways.\cite{Brz07,Brz09_act} We note that the 1D OCM 
is equivalent to the 1D anisotropic XY model, solved exactly in the 
seventies.\cite{Per75} An exact solution of the 1D OCM demonstrates 
that certain NN spin correlation functions change discontinuously at 
the point of a quantum phase transition (QPT) when
both types of interactions have the same strength, similarly to the 2D 
OCM. This somewhat exotic behavior is due to the QPT occurring in this 
case at the multicritical point in the parameter space.\cite{Eri09}
The entanglement measures, together with so called quantum discord in 
the ground state characterizing the quantumness of the correlations,
were analyzed recently\cite{You08,You12} to find the location of 
quantum critical points and to show that the correlations between two 
pseudospins on even bonds are essentially classical in the 1D OCM. 
A slight anisotropy of the interactions leads to particular short-range
correlations dictated by the stronger interaction, but balanced
interactions induce a QPT to a highly degenerate disordered ground
state. 

The 2D OCM is similarly characterized by both classical correlations
on ordered bonds and by large ground state degeneracy. Balanced
interactions $J_x=J_z$ define here a QPT between competing types of 1D 
nematic orders: for $J_x>J_z$ ($J_x<J_z$), antiferromagnetic (AF) 
chains form along $a$ ($b$) that are --- in the thermodynamic limit --- 
not coupled along $b$ ($a$). When going through the QPT, nearest 
neighbor (NN) spin correlations are discontinuous.\cite{Kho03} This 
picture is supported by high-order perturbation theory,\cite{Mil05} a 
rigorous mathematical approach,\cite{You10} mean field (MF) theory on 
the Jordan-Wigner fermions,\cite{Che07} and a sophisticated infinite 
projected entangled-pair state (PEPS) algorithm.\cite{Oru09} 

At the isotropic point $J_x=J_z$, the nematic order with its highly
degenerate ground state manifold persists. It has been shown by quantum 
Monte-Carlo methods to remain stable at finite temperature up to 
$T_{c}=0.055J$ and the phase transition to the fully disordered 
(paramagnetic) phase is in the Ising universality class.\cite{Wen08} 
This resembles periodic frustrated Ising models, where also a phase
transition at finite temperature is found.\cite{Lon80}
The ground state degeneracy was found to be exponential in the linear 
size of the system,\cite{Mil05} implying subextensive entropy at 
zero temperature. As shown by Dou\c{c}ot \textit{et al.},\cite{Dou05} 
the eigenstates of the OCM are twofold degenerate and the number of low 
energy excitations scales as linear size of the system. 
It has also been shown\cite{Cin10} that the isotropic OCM is not 
critical in the sense that the spin waves remain gapful in the ground 
state, confirming that the order in the 2D OCM is not of magnetic type.

While the compass model is used to describe a variety of systems, see
above, it represents a generic simplified concept of the orbital 
physics. Modifications that bring it closer to specific systems have 
been suggested, which allow one on one hand to assess how robust the 
features of the OCM are, and on the other lead to insights about the 
OCM itself. It was proven by exact diagonalization of small 
systems that the low energy excitations of the OCM correspond to the 
spin flips of whole rows or columns of the 2D lattice and that these 
characteristic excitations survive when a small admixture of the 
Heisenberg interactions is included into the compass Hamiltonian.
\cite{Tro10,Tro12} A second generalization interpolates between the 
OCM and an isotropic Ising model, this will be here referred to as the
generalized compass model (GCM). 
The elaborated multiscale entanglement-renormalization ansatz (MERA) 
calculations, together with high-order spin-wave expansion,\cite{Cin10} 
showed that the 2D GCM undergoes a second order QPT between the 
generic OCM and the Ising model.

Recent progress in the 2D OCM was achieved by making use of its 
symmetries. It has been shown that the symmetry allows one to reduce 
the original $L\times L$ compass cluster to a smaller 
$(L-1)\times (L-1)$ one with modified interactions\cite{Brz10} which 
made it possible to obtain exact eigenspectra for larger clusters,
\cite{Brz10_icm,*Brz11_vi3} and investigate the specific heat up to a 
$6\times 6$ system.\cite{Brz13b} 
The spin transformations that provided this reduction were also used to 
uncover a hidden order in the ground state of OCM,\cite{Brz10} 
manifested by the exact identities in the four-spin correlation function 
valid despite imposed anisotropy. It has been shown numerically that 
site dilution reduces ordering temperatures, but keeps the nematic 
character intact.\cite{PhysRevLett.98.256402} Electron itinerancy has 
been addressed in the weak-coupling limit at temperatures above the
ordering transition.~\cite{0295-5075-97-2-27002}

The purpose of this paper is to characterize the motion of a single 
hole in the ordered phases of both the OCM and the GCM, by obtaining the
spectral functions of the itinerant models that reproduce 
both compass models in the strong coupling regime. 
A great advantage of using the itinerant models is that a variational 
cluster approach (VCA) could be used to obtain unbiased results for 
both weak and strong coupling regime. The VCA was introduced to study 
strongly correlated electrons in models with local interactions.
\cite{Dah04,Pot03} 
Recently the VCA was used for the description of the excitonic insulator 
state in the two-orbital Hubbard model,\cite{Kan12} appearing in the 
broad parameter range between band and Mott insulator phases. This
method was successfully applied to investigate hole propagation in the 
$t_{2g}$ orbital model.\cite{Dag08} We will compare its results here
with MF results valid for weak coupling. 

Since superexchange interactions are here 
Ising-like, quantum fluctuations are suppressed and the paradigm for
hole propagation known from the spin $t$-$J$ model, i.e., via coupling
to such fluctuations,\cite{Mar91} may no longer apply. Indeed, it has
been recognized that the Ising-like superexchange arising for $t_{2g}$ 
electrons in $ab$ planes of Sr$_2$VO$_4$ implies that holes move
mostly via three-site terms instead.~\cite{Dag08,*Woh08} In the case of
$e_g$ electrons, describing ferromagnetic (FM) LaMnO$_3$ planes, 
inter-orbital hopping becomes an additional possibility.\cite{vdB00} 

However, all these models show truly 2D magnetic order in the ground 
state. While propagation along the 1D ordered chains of the OCM may be 
expected to show features characteristic of Ising-like order, the second 
and disordered direction presents a qualitatively new challenge. 
Concerning magnetism (resp. orbital superexchange), these bonds are 
inactive in the thermodynamic limit and do not contribute to
the energy. A hole can, in contrast, still hop on these bonds and might
thus in principle mediate couplings between ordered chains. As we are 
going to show, the symmetries of the OCM imply that this does not happen:
The kinetic Hamiltonian of the hole turns out not to depend on the
relative orientation of neighboring chains. We are also going to see 
that propagation in one of the two orbitals reveals the signatures of 
Ising-like order, namely it depends crucially on three-site hopping 
processes allowing for coherent propagation along the ordered chains. 

The paper is organized as follows: In Sec. \ref{sec:hub} we present both 
compass models and their itinerant counterparts. In Sec. \ref{sec:sym} 
we discuss the symmetries specific for the OCM in the context of its 
itinerant version, and in Sec. \ref{sec:mfa} we solve the itinerant 
models in the MF approximation assuming two possible orders of the GCM 
(from Ref. \onlinecite{Cin10}). In Sec. \ref{sec:spec}
we present the VCA spectral functions of the OCM at different $U$,
starting from the weak coupling, metallic regime and ending in the
insulating phase. Finally, in Sec. 
\ref{sec:speg} we present analogical results 
for the GCM at the strong coupling at different values of the control 
parameter $\theta$, starting from the classical limit at $\theta=0$ and 
ending at the critical value $\theta_c$, where the GCM becomes very 
similar to the OCM. Summary and conclusions
are presented in Sec. \ref{sec:summa}. The paper
is supplemented by three appendices with more technical details: 
in Appendix \ref{sec:3site} we show the form of the three-site hopping 
Hamiltonians for the OCM and the GCM, 
in Appendix \ref{sec:basis} we show how the GCM and the OCM can be 
related to each other by the rotation in the space of fermion 
operators at the level of their itinerant models, and 
in Appendix \ref{sec:y} we derive the form of the hopping Hamiltonian 
after the transformation that changes the sign of coupling in the OCM.

\section{Hubbard Hamiltonians for the compass models}
\label{sec:hub}

The quantum compass model (OCM) on a square lattice is defined as
(we consider here AF interactions with $J>0$),
\begin{equation}
{\cal H}_J^0=J\sum_{i}\left\{ \sigma_{i}^{z}\sigma_{i+a}^{z}
+\sigma_{i}^{x}\sigma_{i+b}^{x}\right\},
\label{eq:H_cmp}
\end{equation}
where $\{\sigma_i^x,\sigma_i^z\}$ are $S=1/2$ pseudospin operators and 
$\{i+a(b)\}$ is a shorthand notation for the nearest neighbor of site 
$i$ in the direction $a(b)$. Similarly, the generalized compass model 
(GCM) considered here can be written as
\begin{equation}
{\cal H}_J^{\theta}=J\sum_{i}\left\{ 
 \bar{\sigma}_i( \theta)\bar{\sigma}_{i+a}( \theta)
+\bar{\sigma}_i(-\theta)\bar{\sigma}_{i+b}(-\theta)\right\} ,
\label{eq:H_gcmp}
\end{equation}
where
\begin{eqnarray}
\bar{\sigma}_i( \theta)=
\cos(\theta/2)\sigma_i^x+\sin(\theta/2)\sigma_i^z
\label{eq:sigb}
\end{eqnarray}
are the composed pseudospins interpolating between $\sigma_{i}^{x}$
for $\theta=0$ and $(\sigma_{i}^{x}\pm\sigma_{i}^{z})/\sqrt{2}$ for 
$\theta=\pi/2$. 
For $\theta=0$, this corresponds to the usual Ising model coupling the
$x$ components of spin on all bonds. In the opposite limit
$\theta=\pi/2$, it describes the OCM in a rotated spin space: bonds
along $a$ couple the spin component $S^x +S^z$ and bonds along $b$ the
orthogonal $S^x - S^z$. For $0< \theta < \pi/2$, the GCM interpolates
between Ising and compass models.\cite{Cin10} The rotation of the 
compass model provides an additional convenient way to detect the phase 
transition between 2D-Ising and
nematic compass order: In the former, moments lie along $x$ while they
lie along either $x+z$ (in the following identified with lattice axis
$a$) or $x-z$ in the latter.

Both models can be derived as a large--$U$ limit
of the two-orbital Hubbard model of the form
\begin{eqnarray}
{\cal H}_{t-U} & = & t\sum_{i}\sum_{{\mu,\nu=\atop \alpha,\beta}}\!
\left\{ A_{\mu\nu}c_{i,\mu}^{\dagger}c_{i+a,\nu}^{}
\!+\! B_{\mu\nu}c_{i,\mu}^{\dagger}c_{i+b,\nu}\right\}^{} \!+\! {\rm H.c.}
\nonumber \\
& + & U\sum_{i}n_{i,\alpha}n_{i,\beta},
\label{eq:H_tU}
\end{eqnarray}
at half filling, where $A_{\mu,\nu}$ and $B_{\mu,\nu}$ are hopping
matrices in $a$, $b$ directions between orbitals $\alpha$ and $\beta$.
The hopping matrices
\begin{align}
A_0&=\left(\begin{array}{cc}
1 & 0\\
0 & 0
\end{array}\right) = \frac{1}{2}\,(1+\sigma^z),\\
 B_0&=\frac{1}{2}\left(\begin{array}{cc}
1 & 1\\
1 & 1
\end{array}\right) = \frac{1}{2}\,(1+\sigma^x),
\label{eq:hop_cmp}
\end{align}
for the OCM (\ref{eq:H_cmp}) were given in
Ref.~\onlinecite{0295-5075-97-2-27002}. using standard perturbation 
theory for two neighboring sites one can easily generalize them to the 
GCM Eq. (\ref{eq:H_gcmp}), and one finds that:
\begin{align}
A_{\theta} & =  \frac{1}{\sqrt{2}}\left(\begin{array}{cc}
1+\sin\frac{\theta}{2} & \cos\frac{\theta}{2}\\
\cos\frac{\theta}{2} & 1-\sin\frac{\theta}{2}
\end{array}\right)
=\frac{1}{\sqrt{2}}\bigl[1+\bar{\sigma}(\theta)\bigr],
\label{eq:Ahop_gcmp}\\
B_{\theta} & =  \frac{1}{\sqrt{2}}\left(\begin{array}{cc}
1+\sin\frac{\theta}{2} & -\cos\frac{\theta}{2}\\
-\cos\frac{\theta}{2} & 1-\sin\frac{\theta}{2}
\end{array}\right)
=\frac{1}{\sqrt{2}}\bigl[1-\bar{\sigma}(-\theta)\bigr].
\label{eq:Bhop_gcmp}
\end{align}
The relation between pseudospins $\{\sigma_{i}^x,\sigma_{i}^z\}$ 
and fermions $c_{i}^{\dagger}$ is given by
\begin{equation}
\sigma_{i}^{z}=n_{i,\alpha}-n_{i,\beta},\qquad\sigma_{i}^{x}=
c_{i,\alpha}^{\dagger}
c_{i,\beta}+c_{i,\beta}^{\dagger}c_{i,\alpha},\label{eq:sig_zx}
\end{equation}
and the superexchange constant $J$ is equal to
\begin{equation}
J=\frac{t^{2}}{U}.
\end{equation}
In the large--$U$ limit the $t$-$U$ Hamiltonian can be mapped onto
the $t$-$J$ one. For the compass model this mapping gives,
\begin{equation}
{\cal H}_{t-J}^0={\cal H}_{J}^0+{\cal H}_{t}^0+{\cal H}_{t^{2}}^0,
\end{equation}
with
\begin{equation}
{\cal H}_{t}^0=t\sum_{i}\!\sum_{{\mu,\nu=\atop \alpha,\beta}}\!\!
\left\{\!(A_0)_{\mu\nu}\tilde{c}_{i,\mu}^{\dagger}\tilde{c}_{i+a,\nu}^{}\!
+\!(B_0)_{\mu\nu}\tilde{c}_{i,\mu}^{\dagger}\tilde{c}_{i+b,\nu}\!\right\}^{} 
+  {\rm H.c.}.
\end{equation}
In the tilde fermion operators the double occupancies are projected
out, i.e.,  
$\tilde{c}_{i,\alpha(\beta)}^{\dagger}=c_{i,\alpha(\beta)}^{\dagger}
\left(1-n_{i,\beta(\alpha)}\right)$,
so the $t$-$J$ Hamiltonian at half filling contains no hopping linear
in $t$ and only the $t^{2}$ hopping is possible (three-site hopping).
For the derivation of the three-site hopping ${\cal H}_{t^{2}}^0$
see Appendix \ref{sec:3site}. 

The $t$-$J$ Hamiltonian for the GCM is analogous. As mentioned above, 
the GCM at $\theta=\pi/2$ and OCM are related by the $\pi/4$ rotation 
in the pseudospin space. In the Appendix \ref{sec:basis} we show that 
this implies a similar relation between their fermionic $t$-$U$ 
Hamiltonians.

\section{Symmetries of the orbital compass model}
\label{sec:sym}

The most characteristic symmetries of OCM are the row/column flips
along $x$ or $z$ axis. More precisely, the Hamiltonian ${\cal H}^0_{J}$
of Eq. (\ref{eq:H_cmp}) commutes with $P_{i}$ and $Q_{i}$ operators
defined as,
\begin{equation}
P_{i}=\prod_{n}\sigma_{i+nb}^{z},\qquad Q_{i}=\prod_{n}\sigma_{i+na}^{x}.
\end{equation}
How does it work for the $t$-$U$ compass model Eq. (\ref{eq:H_tU})? The 
operator $Q_i$ should be first generalized to the case of double and 
zero occupancy of site $i$. This can be done by modifying $\sigma_i^x$ 
as follows,
\begin{equation}
\sigma_{i}^{x}\to\tilde{\sigma}_{i}^{x}=
\left(1-n_{i}\right)^{2}+\sigma_{i}^{x},
\end{equation}
so that $(\tilde{\sigma}_{i}^{x})^{2}=1$. Now we can produce new
$\tilde{Q}_{i}$ operator in the same way as before and see its action
on the fermion operators, which is
\begin{equation}
\tilde{Q}_{i}\left(c_{j,\alpha(\beta)}\right)\tilde{Q}_{i}=c_{j,\beta(\alpha)},
\end{equation}
for all $c_{j,\mu}$ lying on the line of $\tilde{Q}_{i}$ and unity
for the others. Under this change the interaction part of the 
${\cal H}_{t-U}^0$ remains unchanged, i.e.,
\begin{equation}
\tilde{Q}_{i}{\cal H}_{U}^0\tilde{Q}_{i}=U\sum_{i}n_{i,\alpha}n_{i,\beta}.
\end{equation}
In the hopping part the hopping matrices $A_0$ and $B_0$
transform by the anti-diagonal transposition, i.e.,
\begin{equation}
A_0=\left(\begin{array}{cc}
1 & 0\\
0 & 0
\end{array}\right)\to\left(\begin{array}{cc}
0 & 0\\
0 & 1
\end{array}\right),
\end{equation}
for the $a$-bonds overlapping with $\tilde{Q}_{i}$ and by unity
for the others. For $b$-bonds incoming to and outgoing from the line
of $\tilde{Q}_{i}$ the same transformation acts as identity,
\begin{equation}
B_0=\frac{1}{2}\left(\begin{array}{cc}
1 & 1\\
1 & 1
\end{array}\right)\to\frac{1}{2}\left(\begin{array}{cc}
1 & 1\\
1 & 1
\end{array}\right),
\end{equation}
so all the \textbf{$b$}-bonds remain unchanged. This brings us to the 
conclusion that ${\cal H}_{t-U}^0$ is covariant under the action of the 
$\tilde{Q}_{i}$; the exact form of the Hamiltonian changes, but the 
change is such that the properties of the new Hamiltonian are the same 
as before --- only the orbitals along one line are renamed which is not 
relevant for the physics. Also the the pseudospin part ${\cal H}_{J}^0$ 
derived out of such a Hamiltonian is the same as before.

We discuss here the AF GCM/OCM model, but it should be noted that all
physical properties remain valid for the FM variants. The equivalence
of FM and AF couplings is of course well known for the Ising limit
$\theta=0$, but since the compass limit $\theta=\pi/2$ is
characterized by frustration, one may wonder whether it is lifted in
the FM variant. This is not the case, as one can see by explicitly
carrying out the transformation. This is done in Appendix
\ref{sec:y} by use of an (anti)symmetry operator that 
anticommutes with ${\cal H}_{J}^0$, which we call $Y$ operator. 
The $Y$ transformation transforms the AF OCM into a FM model, but 
--- as can be seen from the $Y$-transformed hopping in Appendix
\ref{sec:y} --- hole motion remains frustrated in
exactly the same way as in the AF case.

\section{Mean field solution of the generalized compass model}
\label{sec:mfa}

Here we will present a MF solution of the GCM assuming a typical order. 
As the GCM includes OCM as a special case for $\theta=\pi/2$, this
solution will be used later on for both the GCM and its simple version, 
the OCM. As usually in a MF approach, we start from decoupling 
interaction term, i.e., the interaction term of Eq. (\ref{eq:H_tU})
is rewritten as,
\begin{eqnarray}
&\! &n_{i,\alpha}n_{i,\beta}=\frac{1}{2}\left(n_{i,\beta}+n_{i,\alpha}\right)
\nonumber \\
&\!-&\frac{1}{2}\!\left[\cos\!\frac{\varphi}{2}\!
\left(c_{i,\alpha}^{\dagger}c_{i,\beta}\!
+\! c_{i,\beta}^{\dagger}c_{i,\alpha}\right)\!+\!\sin\!\frac{\varphi}{2}\!
\left(n_{i,\alpha}\!-\! n_{i,\beta}\right)\right]^2\!\!,
\end{eqnarray}
where $\varphi$ is an arbitrary angle. In any case we are interested in 
AF type of ordering so the lattice must be divided into two sublattices.
This introduces fermion operators with two flavors defined as follows,
\begin{equation}
\forall i\in A:\hskip .3cm c_{i,\mu}^{\dagger}=c_{i,\mu}^{A\dagger},
\quad c_{i+a,\mu}^{\dagger}=c_{i,\mu}^{B\dagger}.
\end{equation}
Now we introduce mean field $h$ which interpolates between $\sigma^{x}$
or $\sigma^{z}$ magnetization depending on $\varphi$,
\begin{equation}
h\equiv\left\langle\cos\!\frac{\varphi}{2}\!
\left(c_{i,\alpha}^{B\dagger}c_{i,\beta}^{B}\!
+\! c_{i,\beta}^{B\dagger}c_{i,\alpha}^{B}\right)\!
+\!\sin\!\frac{\varphi}{2}\!
\left(n_{i,\alpha}^{B}\!-\! n_{i,\beta}^{B}\right)\!\right\rangle .
\end{equation}
Using the above equation we are ready to write the $t$-$U$ Hamiltonian
of Eq. (\ref{eq:H_tU}) in a MF form in the $\vec{k}$-space,
\begin{eqnarray}
& &H^{\rm MF}_{t-U}=2t\sum_{k\in A}\sum_{{\mu,\nu=
\atop \alpha,\beta}}\!\gamma_{\vec{k}}^{\mu\nu}
c_{\vec{k},\mu}^{A\dagger}c_{\vec{k},\nu}^{B}
+ {\rm H.c.}\nonumber \\
&-&\frac{U}{2}h\!\sum_{\vec{k}\in A}\!\!\left[\cos\!\frac{\varphi}{2}\!
\left(c_{\vec{k},\alpha}^{A\dagger}
c_{\vec{k},\beta}^{A}\!+\! c_{\vec{k},\beta}^{A\dagger}
c_{\vec{k},\alpha}^{A}\!\right)\!
+\!\sin\!\frac{\varphi}{2}\!\left(n_{\vec{k},\alpha}^{A}\!
-\! n_{\vec{k},\beta}^{A}\right)\!\right] \nonumber \\
&+&\frac{U}{2}h\!\sum_{\vec{k}\in A}\!\!\left[\cos\!\frac{\varphi}{2}\!
\left(c_{\vec{k},\alpha}^{B\dagger}
c_{\vec{k},\beta}^{B}\!+\! c_{\vec{k},\beta}^{B\dagger}
c_{\vec{k},\alpha}^{B}\!\right)\!
+\!\sin\!\frac{\varphi}{2}\!\left(n_{\vec{k},\alpha}^{B}\!
-\! n_{\vec{k},\beta}^{B}\right)\!\right] \nonumber \\
&+&\frac{U}{2}\sum_{\vec{k}\in A}\left(n_{\vec{k},\alpha}^{A}
+n_{\vec{k},\beta}^{A}+n_{\vec{k},\alpha}^{B}
+n_{\vec{k},\beta}^{B}\right),
\label{eq:t-U_mf}
\end{eqnarray}
where
\begin{equation}
\gamma_{\vec{k}}^{\mu\nu}\equiv A_{\mu,\nu}\cos k_{a}\!+\! B_{\mu,\nu}\cos k_{b}.
\end{equation}
The last step is Bogoliubov transformation. We introduce new fermion
operators $f_{k,\mu}^{S\dagger}$ for $S=A,B$ and $\mu=\alpha,\beta$
being linear combination of the old ones,
\begin{equation}
f_{\vec{k},\mu}^{S\dagger}={\cal L}\left(c_{\vec{k},\alpha}^{A\dagger},
c_{\vec{k},\beta}^{A\dagger},c_{\vec{k},\alpha}^{B\dagger},
c_{\vec{k},\beta}^{B\dagger}\right).
\end{equation}
The eigenmodes can be determined by the equation,
\begin{equation}
\left[H^{\rm MF}_{t-U},f_{\vec{k},\mu}^{S\dagger}\right]=
E_{\vec{k},\mu}^{S}f_{\vec{k},\mu}^{S\dagger}.
\end{equation}
Thus the transformation matrix ${\cal B}$ reads,
\begin{equation}
{\cal B}\!=\frac{U}{2}{\bf 1}+\frac{U}{2}\!\left(\begin{array}{cccc}
-h\sin\frac{\varphi}{2} & -h\cos\frac{\varphi}{2} & \frac{4t}{U}
\gamma_{\vec{k}}^{\alpha\alpha} & \frac{4t}{U}\gamma_{\vec{k}}^{\alpha\beta}\\
-h\cos\frac{\varphi}{2} & h\sin\frac{\varphi}{2} & \frac{4t}{U}
\gamma_{\vec{k}}^{\beta\alpha} & \frac{4t}{U}\gamma_{\vec{k}}^{\beta\beta}\\
\frac{4t}{U}\gamma_{\vec{k}}^{\alpha\alpha} & \frac{4t}{U}
\gamma_{\vec{k}}^{\beta\alpha} & h\sin\frac{\varphi}{2} & h\cos\frac{\varphi}{2}\\
\frac{4t}{U}\gamma_{\vec{k}}^{\alpha\beta} & \frac{4t}{U}
\gamma_{\vec{k}}^{\beta\beta} & h\cos\frac{\varphi}{2} & -h\sin\frac{\varphi}{2}
\end{array}\right)\!.
\end{equation}
After diagonalization of ${\cal B}$ we get four eigenenergies two
of which are smaller than the others --- we denote them as 
$\{E_{\vec{k},1}^{<},E_{\vec{k},2}^{<}\}$.
After filling the system with one fermion per site we obtain the ground 
state energy per site, ${\cal E}_{0}$, as an integral over the reduced
Brillouin zone, i.e.,
\begin{equation}
{\cal E}_{0}=\frac{1}{8\pi^{2}}\int_{-\pi}^{\pi}dk_{u}dk_{v}\left\{ E_{\vec{k},1}^{<}+E_{\vec{k},2}^{<}\right\} ,
\end{equation}
with $k_{a}=(k_{u}+k_{v})/2$ and $k_{b}=(k_{u}-k_{v})/2$. 
The self-consistency equation of the form,
\begin{equation}
\frac{2}{U}\frac{d}{dh}{\cal E}_{0}=h,
\end{equation}
can be solved numerically by performing 
the numerical integration in ${\cal E}_{0}$.

\begin{figure}[t!]
\includegraphics[clip,width=8cm]{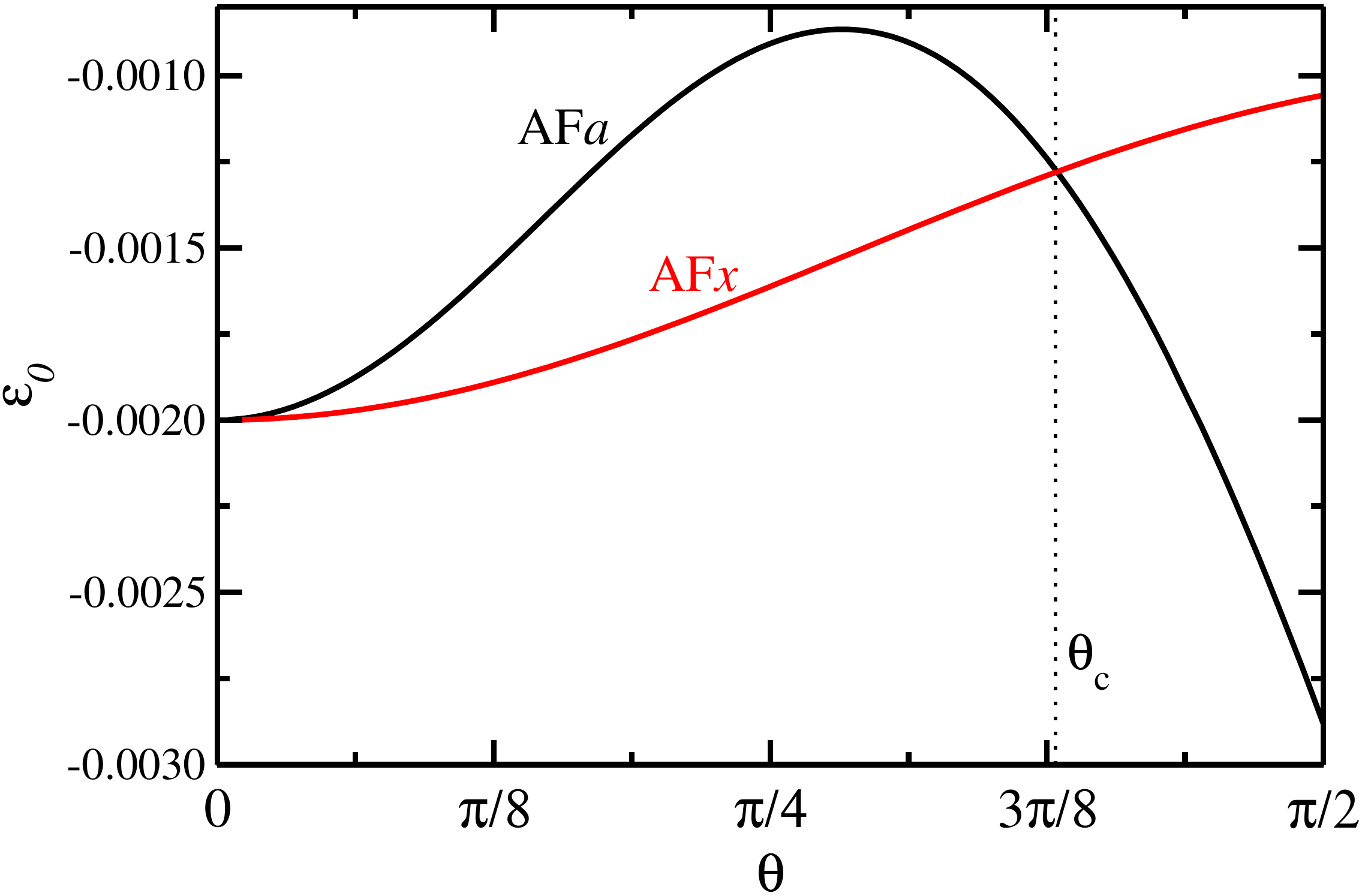}
\caption{
Ground state energies in the AF$x$ and AF$a$ phases of the GCM
at $U=20t$ as functions of angle $\theta$ obtains in the mean field.
Critical angle $\theta_{c}^{\rm MF}$ is marked with dotted line, here
$\theta_{c}^{\rm MF}\approx 68^{\circ}$.
\label{fig:Ground-state-energies}}
\end{figure}

Following the results for the generalized compass model presented
in Ref. \onlinecite{Cin10} we impose two different orderings depending
on the angle $\theta$, entering hopping matrices $\{A,B\}$ as shown by 
Eqs. (\ref{eq:Ahop_gcmp}), (\ref{eq:Bhop_gcmp}). The first order occurs 
below the critical angle $\theta_{c}^{\rm MERA}\approx84.8^{\circ}$
(according to the MERA results of Ref. \onlinecite{Cin10}) and this is
AF order in the $\sigma^{x}$ components of the pseudospins (AF$x$),
thus we take MF $h$ with $\varphi=\pi$ to simulate this phase. Above 
$\theta_{c}$ the order changes into AF order that tracks one of the 
effective pseudospins $\bar{\sigma}(\theta)$ or $\bar{\sigma}(-\theta)$ 
(AF$a$) of Eq. (\ref{eq:sigb}), so we set $\varphi=\theta$ in 
the definition of $h$. In Fig. \ref{fig:Ground-state-energies}
we show the energies ${\cal E}_{0}$ for these two phases as functions
of $\theta$ at $U=20t$. We can see that their behavior is qualitatively
correct, i.e., for $\theta=0$ the two phases are the same, so the
energies are equal. When $\theta$ increases, the AF$x$ becomes 
favorable until $\theta=\theta_{c}^{\rm MF}$ and in our case 
$\theta_{c}^{\rm MF}\approx 68^{\circ}$.
Above $\theta_{c}$ the AF$a$ phase is favorable.

\section{Spectral functions of the orbital compass model at different $U$}
\label{sec:spec}

In order to address the strong-coupling limit at large $U$, where the
$t$-$U$ models come close to the OCM and GCM models, we use the VCA. 
It builds on cluster-perturbation theory, where the self-energy is 
calculated exactly (using exact diagonalization) for a small cluster 
and used to evaluate the one-particle Green's function of a much larger 
system. This is complemented by optimization, where the grand potential
is minimized with respect to a proposed order parameter. However, the
method is only applicable to order parameters that are quadratic in 
fermion operators, e.g. magnetic, orbital or superconducting order. 
The order parameter for a nematic phase, in contrast,  
is proportional to $\left\langle \sigma_{i}^{z}\sigma_{i+a}^{z}
-\sigma_{i}^{x}\sigma_{i+b}^{x}\right\rangle $, and the VCA can
consequently not be used to self-consistently detect this order in the
GCM.\cite{Daghofer:2012kl} Nevertheless, cluster-perturbation theory
has been shown to be useful in obtaining spectral function for a 
nematic ground state imposed on models for iron-based superconductors.
\cite{Daghofer:2012ch} 

\begin{figure}[t!]
\begin{minipage}[t]{1\columnwidth}%
\begin{center}\includegraphics[clip,width=8cm]{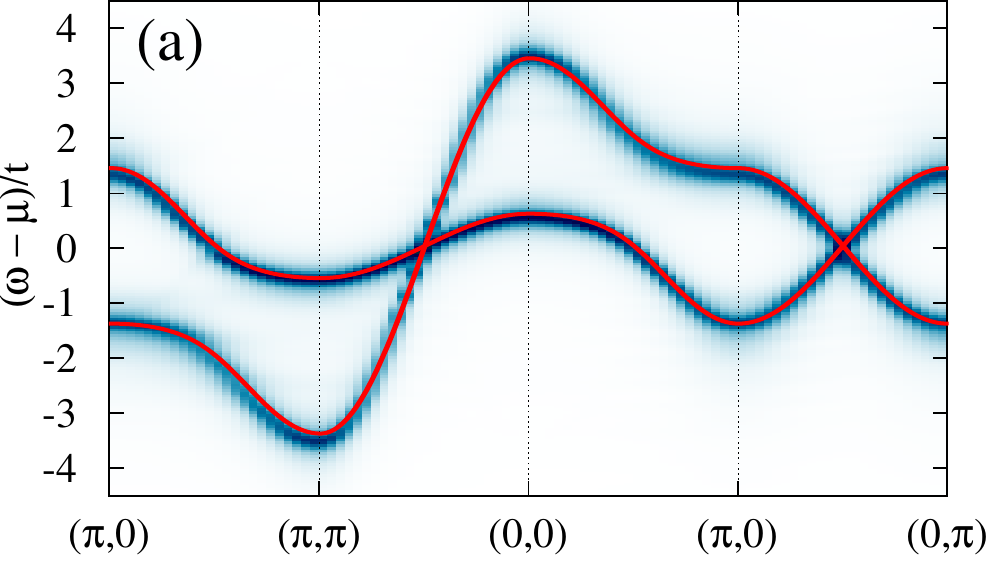}\end{center}%
\end{minipage}
\begin{minipage}[t]{1\columnwidth}%
\begin{center}\includegraphics[clip,width=8cm]{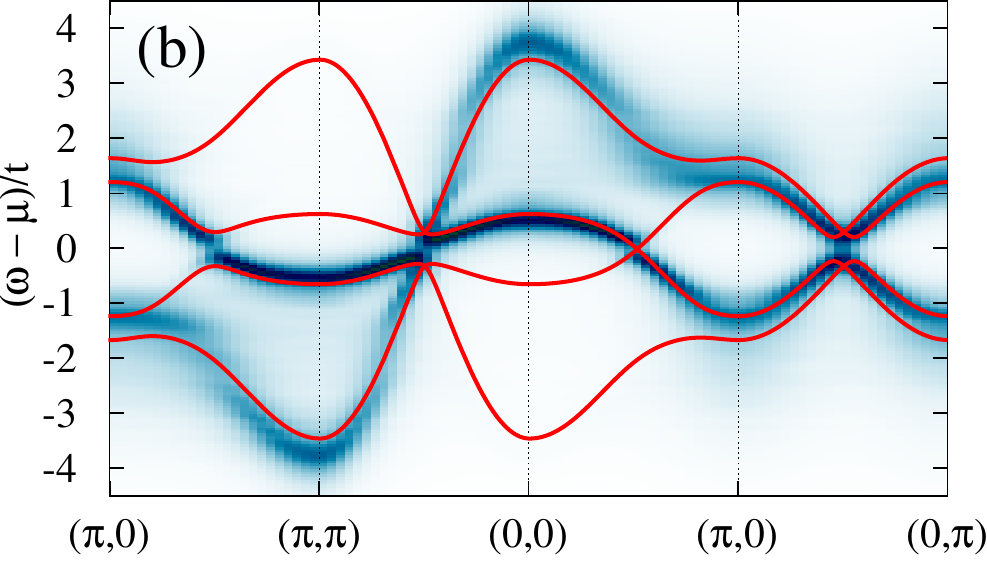}\end{center}%
\end{minipage}
\begin{minipage}[t]{1\columnwidth}%
\begin{center}\includegraphics[clip,width=8cm]{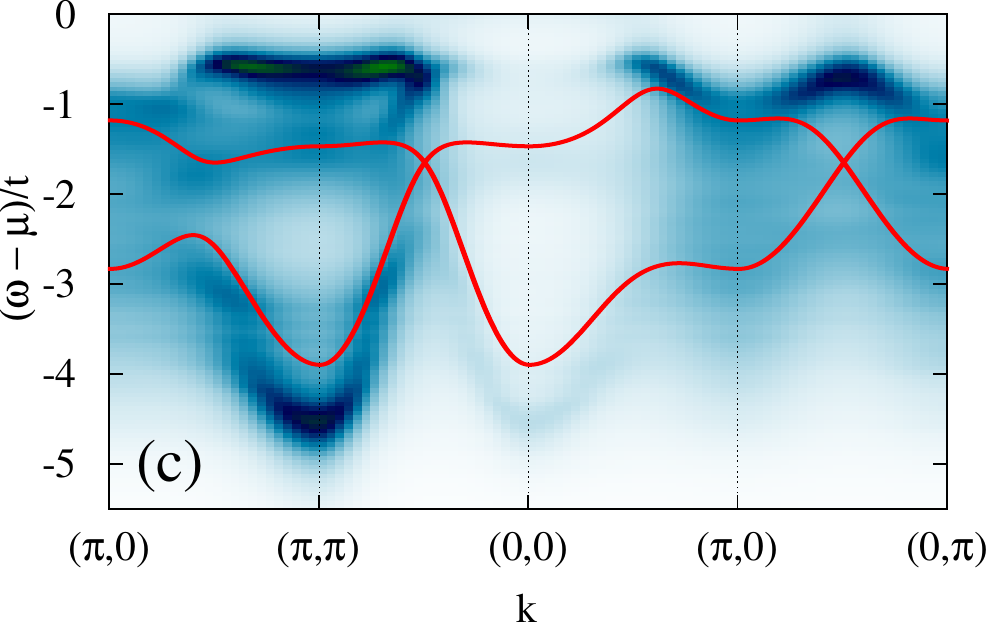}\end{center}%
\end{minipage}
\caption{Spectral functions for the OCM found at: 
(a) $U=1t$, 
(b) $U=2t$, and 
(c) $U=4t$. 
In the last case only the hole-part of the spectral
function is shown. Solid lines stand for the MF bands. }
\label{fig:spec_cmp_U}
\end{figure}

Here, where the microscopic character of the nematic state is slightly
different, we use a different approach. We make use of the facts that:
(i) the nematic state is (in the thermodynamic limit) given by
mutually decoupled AF chains, and 
(ii) the $t$-$U$ Hamiltonian for the hole does not depend on the mutual 
orientation of the chains, see Sec.~\ref{sec:sym}. 
To treat the nematic state, we thus set the order parameter to select 
one configuration out of the ground state manifold, e.g. the AF one. 
The grand potential can be optimized just as in the AF state and the 
spectral density can be obtained, which is identical to that of all 
other ground states. This approach neglects tunneling from one nematic 
state to any other, but as the time scale of the related flip of a 
whole chain is much longer than the time scale of hole motion, 
especially in large systems, this is not expected to affect the hole's 
motion.

We present here the VCA results obtained for a directly solved cluster 
of $3\times4$ sites, with superlattice translation vectors, being 
$\vec{x}=\left(3,1\right)$ and $\vec{y}=\left(0,4\right)$, so that the 
AF order within the cluster implies that the whole lattice is AF.
We also used other cluster geometries and sizes, e.g. 
$\sqrt{10}\times\sqrt{10}$, for comparison and found consistent
results, suggesting that the features that we observe in spectral 
functions are not cluster-dependent.

\begin{figure}[t!]
\begin{minipage}[t]{1\columnwidth}%
\begin{center}\includegraphics[clip,width=8cm]{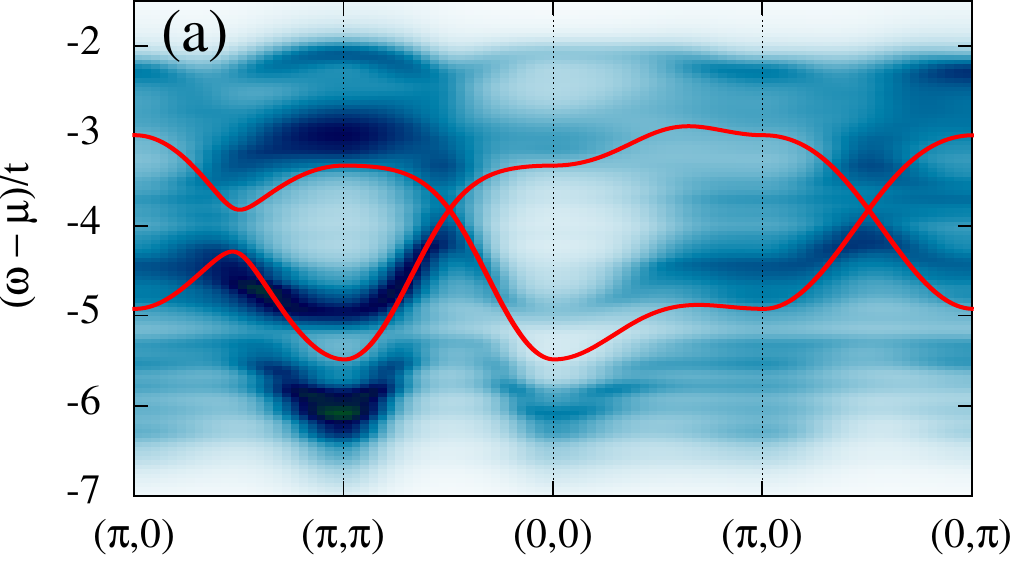}\end{center}%
\end{minipage}
\begin{minipage}[t]{1\columnwidth}%
\begin{center}\includegraphics[clip,width=8cm]{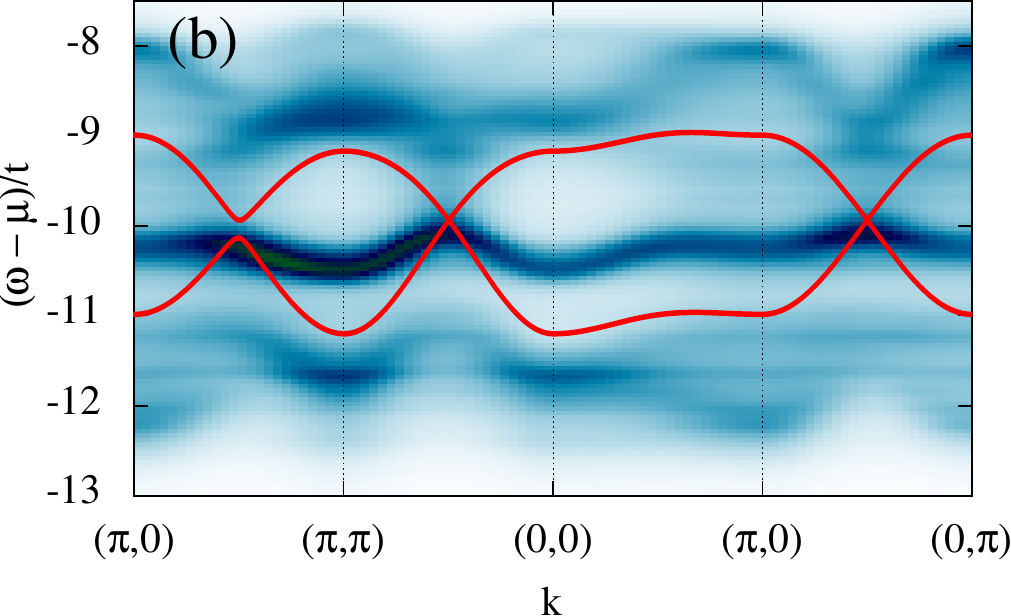}\end{center}%
\end{minipage}
\caption{Spectral functions for the OCM found at: 
(a) $U=8t$, and 
(b) $U=20t$.
Only the hole part is shown. Solid lines stand for the MF bands. }
\label{fig:spec_cmp_U2}
\end{figure}

In Fig. \ref{fig:spec_cmp_U}, we present the VCA spectral functions in 
the limit of small $U$ together
with the MF bands.  Figure \ref{fig:spec_cmp_U}(a)
shows the spectral function for $U=1t$ along a standard path in the 
Brillouin zone. In this weak-interaction limit, the VCA spectral
function exhibits two coherent bands coinciding with the bands obtained
in the MF approach. This confirms the correctness of the numerical
treatment. The dispersion of the bands shows the high mobility of the 
hole especially around the point $\vec{k}=\left(\pi/2,\pi/2\right)$,
and it is clearly visible that one band is more dispersive than the
other. For $U=2t$, see Fig. \ref{fig:spec_cmp_U}(b), the system has 
already gone through a metal-insulator transition in the MF approach 
and two subbands have formed that turn out to correspond to the upper
and lower Hubbard bands. In the VCA, the system is indeed 
close to the transition, so that some spectral weight is transferred to 
the new (shadow) bands, but this weight is very small so that the bands 
are not yet visible in the plot. The coherent part of the VCA spectral 
function still coincides with the MF bands, but some incoherent features 
can already be recognized. Further increase in $U$ up to $U=4t$ is 
enough to drive the cluster through the metal-insulator transition and 
to split the hole and electron parts of the
spectral function in the VCA --- in Fig. \ref{fig:spec_cmp_U}(c) we show
the hole spectral function for $U=4t$, i.e., the lower Hubbard band. 
Except for some momentum-space regions around
$\vec{k}=\left(0,0\right)$, interactions now induce more
incoherence. Nevertheless, coherent bands can be clearly observed, see
e.g. the sharp features in the lowest and highest energy part of the
spectral function around $\vec{k}=\left(\pi,\pi\right)$. Qualitatively
the VCA and MF bands look quite similar on the path between
$\vec{k}=\left(\pi,\pi\right)$ and $\vec{k}=\left(\pi,0\right)$. This
is not the case for the section between $\vec{k}=\left(\pi,0\right)$
and $\vec{k}=\left(0,\pi\right)$, where the sharp feature around 
$\vec{k}=\left(\pi/2,\pi/2\right)$ has opposite convexity than the MF 
band. Finally, around $\vec{k}=\left(\pi,0\right)$ the VCA spectral 
function does not exhibit any coherent features, but has only incoherent 
spectral weight.

\begin{figure}[t!]
\begin{minipage}[t]{1\columnwidth}%
\begin{center}\includegraphics[clip,width=8.2cm]{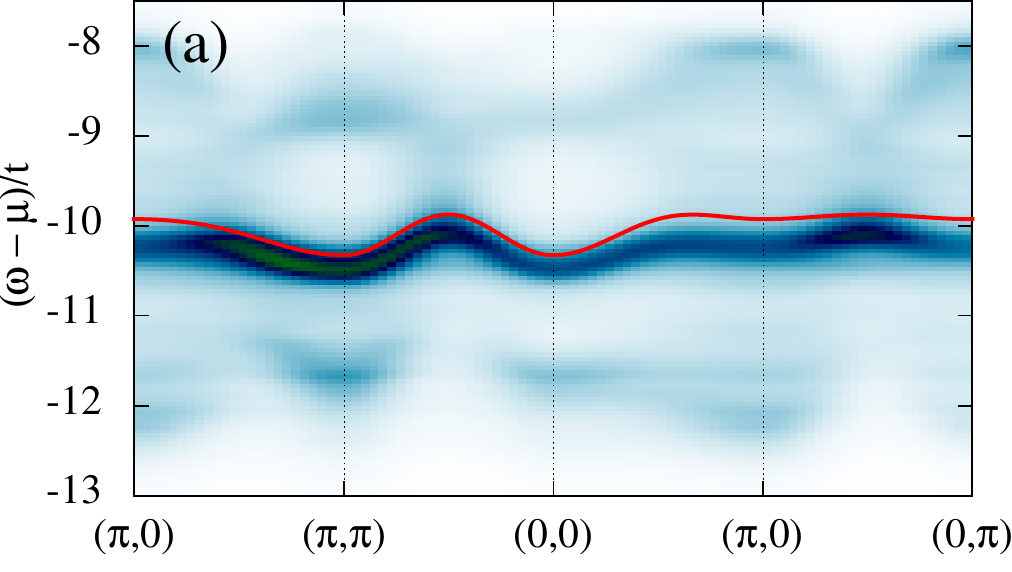}\end{center}%
\end{minipage}
\begin{minipage}[t]{1\columnwidth}%
\begin{center}\includegraphics[clip,width=8.2cm]{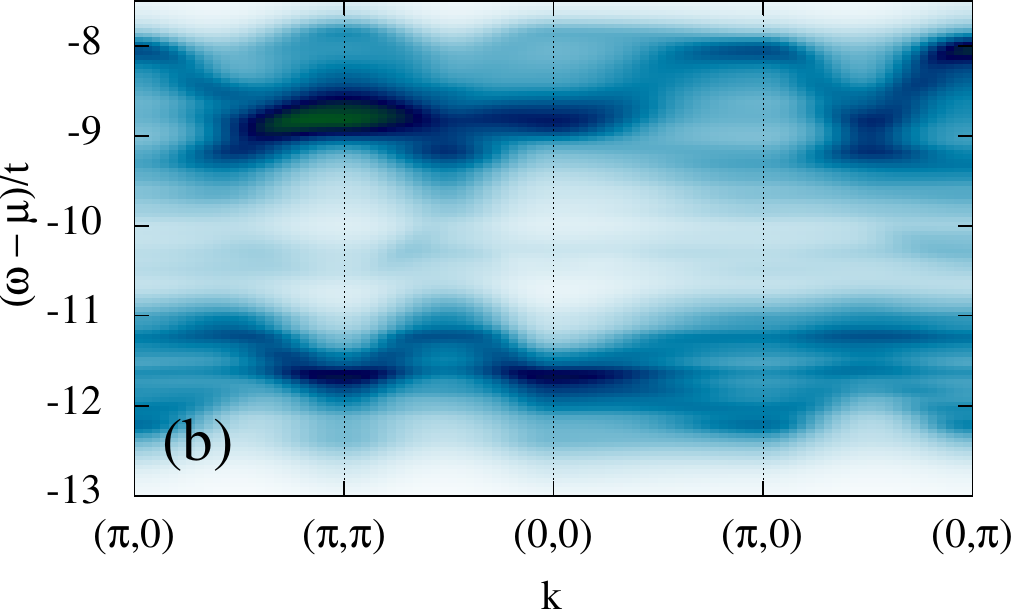}\end{center}%
\end{minipage}
\caption{Spectral functions for the OCM at $U=20t$ projected on 
a single orbital, as obtained for:
(a) $\alpha$ orbitals and 
(b) $\beta$ orbitals. 
The solid line is a MF three-site hopping band for $\alpha$ orbitals. }
\label{fig:spec_cmp_orbs}
\end{figure}

Figure \ref{fig:spec_cmp_U2} shows the spectral functions in the limit 
of large $U$. For $U=8t$, see Fig. \ref{fig:spec_cmp_U2}(a), we can see 
that the spectral weight is distributed more equally among the states 
around $\vec{k}=\left(0,0\right)$ and $\vec{k}=\left(\pi,\pi\right)$ 
than for smaller values of $U$. The bottom band is seen as a coherent 
feature roughly agreeing with the MF prediction, but much less 
dispersive. The upper band cannot be identified easily with any MF band, 
even though although the features around
$\vec{k}=\left(\pi/2,\pi/2\right)$ resemble the MF bands. Especially in 
the large-$U$ limit, see  Fig. \ref{fig:spec_cmp_U2}(b) for $U=20t$, the 
weight imbalance between $\vec{k}=\left(0,0\right)$ and 
$\vec{k}=\left(\pi,\pi\right)$ is no longer visible and the bands are
flatter than for lower $U$, both in qualitative agreement with the MF
results. Also the shapes of the bands in VCA agree to some extent with 
the MF bands, especially around $\vec{k}=\left(\pi/2,\pi/2\right)$. 
Strong coupling differences to the MF bands are on one hand the 
incoherent weight and on the other the separation of bottom and top 
bands. Even the MF bands do not really cross, but they 
remain very close to each other at $\vec{k}=\left(\pi/2,\pi/2\right)$. 
In the VCA, the are much further separated, which means there is a 
strong effective interaction at this value of $\vec{k}$ that cannot 
be captured by a simple MF approach.

Maybe the most obvious new feature seen at large $U$ is, however, a
rather coherent band in the middle of the spectrum. It has strongest 
intensity around $\vec{k}=\left(\pi,\pi\right)$ and can be best seen in 
Fig. \ref{fig:spec_cmp_U2}(b) for $U=20t$, where it is the sharpest 
feature of the spectral function. The extra band is absent from the MF 
approach and in the VCA, it seems to strongly repel the two bands at 
the top and bottom of the spectrum, thus making them flatter and the 
overall spectrum much wider than in the MF approach.  

To better understand the results in the strong coupling regime we
have projected the spectral function on $\alpha$ and $\beta$ orbitals
for $U=20t$. This is shown in Fig. \ref{fig:spec_cmp_orbs}(a) and
\ref{fig:spec_cmp_orbs}(b). Comparing Fig. \ref{fig:spec_cmp_orbs}(a)
with the initial, nonprojected, result of Fig. \ref{fig:spec_cmp_U2}(b)
we can see that almost only the central band is visible in the $\alpha$ 
channel and is very sharp. The $\beta$-projection in Fig. 
\ref{fig:spec_cmp_U2}(b) conversely only shows the top and bottom bands. 
The central band absent from MF spectra can now be identified as due to
three-site hopping of $\alpha$ orbitals along the AF chains. Noting 
this and assuming classical AF order in the OCM ground state, we can 
easily derive an approximate form of the three-site hopping band from 
the general three-site hopping Hamiltonian of Eq. (\ref{eq:cmp_3site})
by putting $\tilde{n}_{i,\alpha}=0$ and $\tilde{n}_{i,\beta}=1$ for 
$i\in A$ and thus $\tilde{n}_{i,\alpha}=1$ and $\tilde{n}_{i,\beta}=0$
for $i\in B$ in the central site of the hopping term. This leads to the 
kinetic Hamiltonian of the form,
\begin{eqnarray}
H_{t^{2}}^{\alpha} & = & -\frac{2t^{2}}{U}\sum_{k\in A}
\tilde{c}_{k,\alpha}^{B\dagger}\tilde{c}_{k,\alpha}^{B}
\left\{ \cos\left(2k_a\right)+\frac{1}{4}\cos\left(2k_b\right)\right.
\nonumber \\
&+& \left.\cos\left(k_{a}+k_{b}\right)+\cos\left(k_a-k_b\right)\right\}.
\label{eq:3s_alph_app}
\end{eqnarray}
This dispersion relation is shown in Fig. \ref{fig:spec_cmp_orbs}(a)
and indeed reproduces well the band obtained by the VCA.

\begin{figure}[t!]
\begin{minipage}[t]{1\columnwidth}%
\begin{center}\includegraphics[clip,width=7cm]{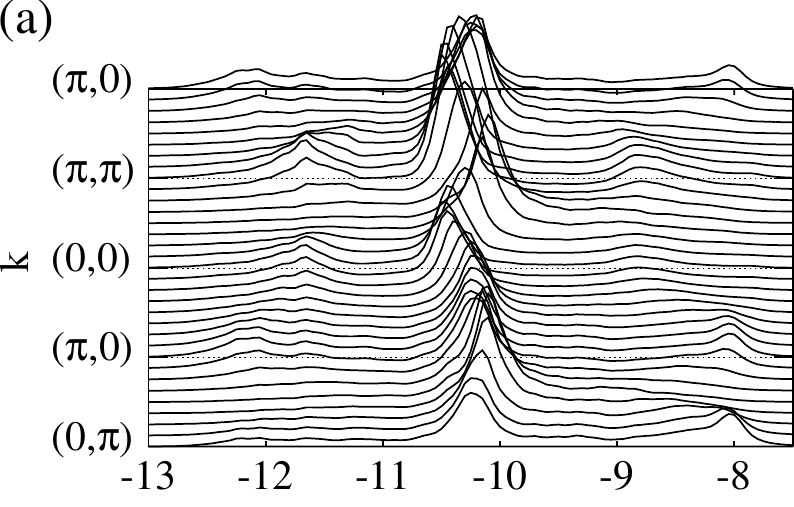}\end{center}%
\end{minipage}
\begin{minipage}[t]{1\columnwidth}%
\begin{center}\includegraphics[clip,width=7cm]{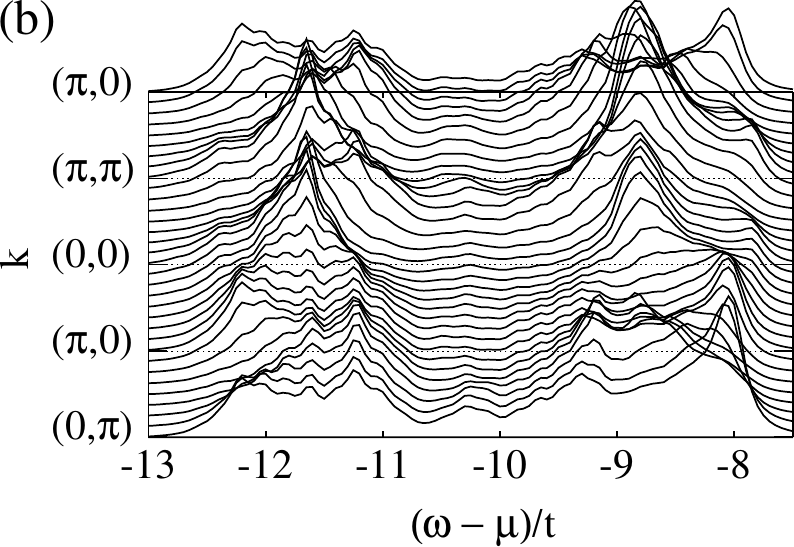}\end{center}%
\end{minipage}
\caption{Spectral functions for the OCM at $U=20t$ shown as a 3D fence 
plot, projected on a single orbital, as obtained for:
(a) $\alpha$ orbital states, and 
(b) $\beta$ orbitals. }
\label{fig:spec_cmp_orb_3d}
\end{figure}

Alternative plots of the projected spectral functions of Figs.
\ref{fig:spec_cmp_orbs} are presented in Figs. 
\ref{fig:spec_cmp_orb_3d}. Here the 3D fence plots are used instead of 
the map plots. As before, the three-site hopping band is well visible
in Fig. \ref{fig:spec_cmp_orbs}(a) as a ridge of tall, coherent peaks
and the other features can be seen in Fig. \ref{fig:spec_cmp_orbs}(b).
Most of them are incoherent.

\section{Spectral functions of the generalized compass model}
\label{sec:speg}
 
In this Section we present spectral functions of the GCM obtained via 
the VCA for a few selected values of angle $\theta$. 
We have used the same procedure as for the original
OCM: the $t$-$U$ Hamiltonian (\ref{eq:H_tU}) for the GCM was 
implemented into a VCA input file and the optimization of the grand
potential $\Omega$ was done with respect to the order parameter.
Following the results from Ref. \onlinecite{Cin10} and Sec. 
\ref{sec:mfa}, we assumed two possible orders, AF$a$ and AF$x$ one, and 
we compared the optimal values of $\Omega$ for each of them to decide 
which configuration is more favorable for a selected value of $\theta$. 
We have verified that the VCA results for lower values of $U\lesssim 8t$ 
show a preference for the AF$x$ direction for all values of $\theta$. 
For $\theta\to\pi/2$, this is in contrast to the expectations (and to 
the VCA results) for the strong coupling limit, where the model goes 
over into the OCM and prefers AF$a$ order for 
$\theta>\theta_{c}^{\textrm{VCA}}\approx88^{\circ}$. We note here that 
the ground state manifold of the classical OCM model has in fact an
accidental degeneracy that makes AF$x$ and AF$a$ (as well as orientation 
along any other direction) equivalent and which is only lifted by 
quantum and thermal fluctuations.\cite{Mish04} In the case of the
itinerant model, orbital fluctuations in the weak-coupling regime have
been noted to differ from strong coupling, which has been attributed to 
a different degree of band hybridization.\cite{0295-5075-97-2-27002} 
The importance of $U$-dependent hybridization and the close energies of 
various orientations are probably the reason for the basis-sensitivity 
of the VCA results. As we are here interested in hole dynamics of the 
OCM and GCM, we focus on larger values of $U$, where all results are 
consistent.

\begin{figure}[t!]
\begin{minipage}[t]{1\columnwidth}%
\begin{center}\includegraphics[clip,width=8cm]{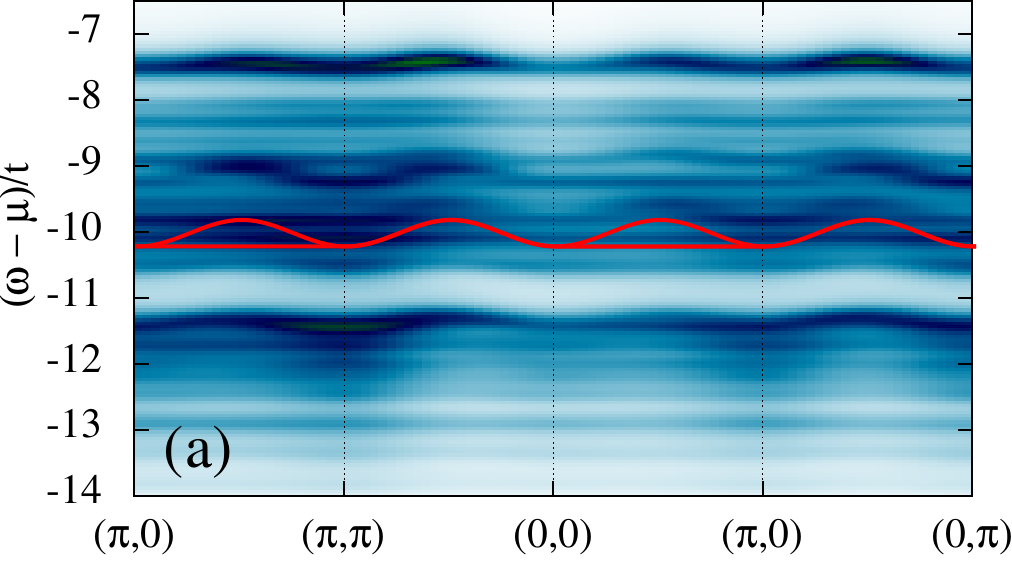}\end{center}%
\end{minipage}
\begin{minipage}[t]{1\columnwidth}%
\begin{center}\includegraphics[clip,width=8cm]{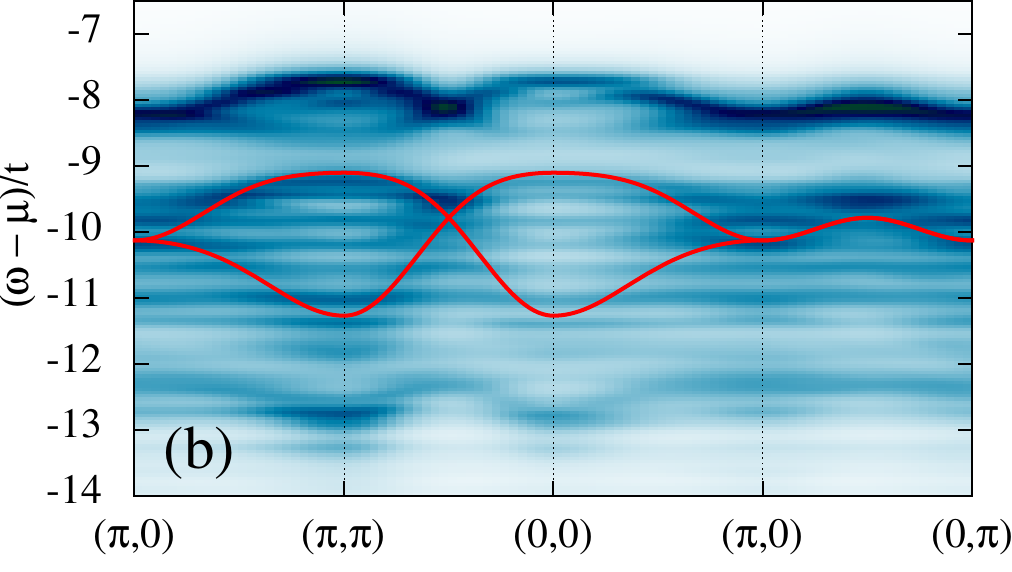}\end{center}%
\end{minipage}
\begin{minipage}[t]{1\columnwidth}%
\begin{center}\includegraphics[clip,width=8cm]{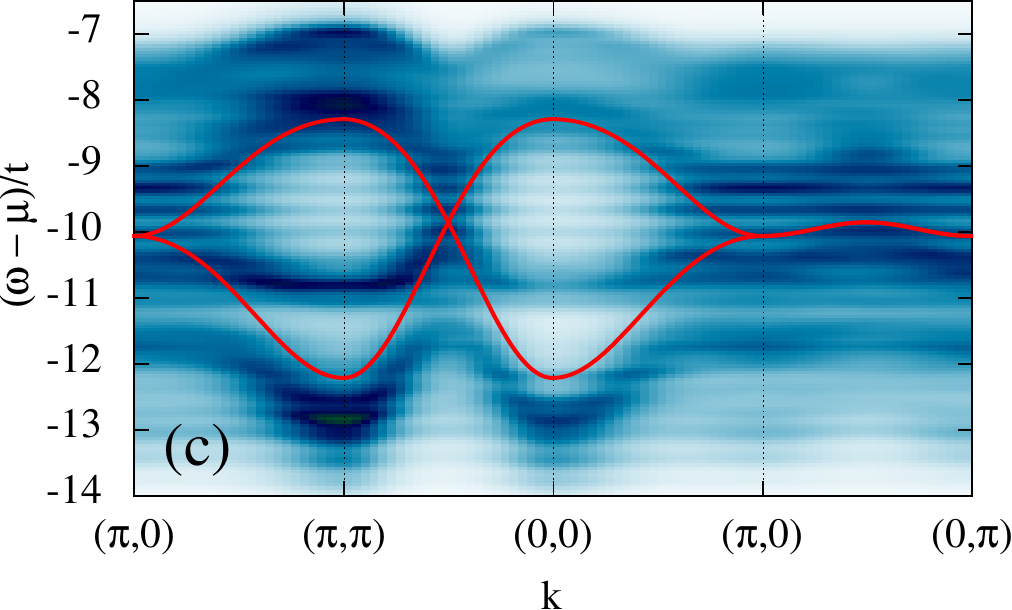}\end{center}%
\end{minipage}
\caption{
Spectral functions obtained at strong coupling for the AF$x$ phase in 
the GCM for increasing frustration of interactions, given by:
(a) $\theta=0$,
(b) $\theta=\pi/4$, and 
(c) $\theta=88^{\circ}-\epsilon$. 
Solid lines stand for the MF bands. 
Parameter: $U=20t$.}
\label{fig:spec_gcmp}
\end{figure}

We are here going to analyze how the spectral density of the 2D Ising
magnet with AF$x$ order evolves when going from the pure Ising model
towards the transition to nematic order, i.e., for increasing $\theta$.  
In Figs. \ref{fig:spec_gcmp}, we show the VCA spectral functions of GCM 
at three different angles $\theta<\theta_{c}$ for $U=20t$.
As for OCM, the results were tested for finite-size effects by changing
cluster geometry and size; results presented here are for a $3\times4$
cluster. Figure~\ref{fig:spec_gcmp}(a) shows the hole spectral function
for $\theta=0$, where the GCM reduces to the classical AF Ising model.
The overall spectrum has approximately ladder character, as expected,
because the hole is confined in a string potential and quantum
fluctuations which might relieve the confinement are absent. The only
mechanism allowing for weak dispersion is three-site hopping, which 
acts on the scale of $t^2/U$.\cite{Dag08} The two MF bands cannot of 
course reflect the ladder spectrum, i.e., both energies and total width 
of the spectrum are wrong, but they do reflect the low hole mobility. 

For $\theta=\pi/4$, see Fig. \ref{fig:spec_gcmp}(b), the bands become 
significantly more dispersive, especially the ones on the top,
while the ones on the bottom are less dispersive. The shape of the
topmost band is qualitatively well reproduced by the MF and this band
is the sharpest feature seen in the spectral function at $\theta=\pi/4$.
As in the case of the original OCM at high $U$ 
(Fig. \ref{fig:spec_cmp_U2}) the bands predicted by MF repel each other 
in the VCA and new features emerge at the intermediate energies, 
with rather incoherent weight.
Similarly to the generic OCM case, bands are most dispersive along the 
direction $(0,0) \to (\pi,\pi)$. The increased dispersion, especially 
of the rather coherent topmost band, is here not primarily driven by 
quantum fluctuations, because the ground state is still Ising ordered, 
as it is fond for $\theta=0$, see above. However, interorbital hopping 
is now active, see Eqs.~(\ref{eq:Ahop_gcmp}) and (\ref{eq:Bhop_gcmp}), 
which allows the hole to evade the string potential and to propagate, 
similar to the case of a hole in $e_g$ orbital order.\cite{vdB00}

Finally, in Fig. \ref{fig:spec_gcmp}(c), we show the spectral function 
infinitesimally close to the transition angle $\theta_{c}$, thus 
$\theta=88^{\circ}-\epsilon$. As in the MF results the bands are 
more dispersive and the agreement between both approximations is better. 
Comparing to $\theta=\pi/4$, spectral weight is distributed more equally 
on the energy scale and it shows imbalance between 
$k=\left(\pi,\pi\right)$ and $k=\left(0,0\right)$,
similar to the OCM at $U=8t$, see Fig. \ref{fig:spec_cmp_U2}(a).
Despite a relatively large value of $\theta$ the overall ladder 
modulation of the spectrum, characteristic for the Ising model at 
$\theta=0$, is still well visible. This 
is a consequence of small quantum fluctuations in the ground state of 
the undoped GCM, as shown in Ref. \onlinecite{Cin10}.
Therefore we should attribute all the difference in hole's behavior 
induced by growing $\theta$ rather to its $\theta$-dependent hopping 
term than to the change of its background. Note that the inter-orbital 
hopping of the hole may induce quantum fluctuations as well, but they 
should be distinguished from fluctuations inherent in the undoped 
ground state. On the other hand, for values of $\theta$ higher than 
$\theta_{c}^{\textrm{\rm VCA}}$ the system is already in
the AF$a$ phase and its spectral function is very similar to the
one already discussed in case of the OCM.

\section{Summary and Conclusions\label{sec:summa}}

We have derived an itinerant $t$-$U$ model for the generic as well as 
for generalized compass models
by choosing proper hopping amplitudes of the respective spinless 
two-band Hubbard models. The itinerant models studied here reproduce 
the form of the generalized (and generic orbital) compass model in the 
limit of large interaction $U$, when electrons localize and orbital 
degrees of freedom are coupled by the superexchange processes. The 
$t$-$U$ models were then solved in the mean field approximation by 
splitting the interaction term and assuming antiferromagnetic order 
of checkerboard type (either AF$a$ or AF$x$ type). The mean field 
approach predicted correctly the transition between AF$a$ and AF$x$ 
order in the generalized compass model, however, the critical angle 
$\theta_{c}^{\rm MF}\approx 68^{\circ}$ (at strong coupling, $U=20t$) is 
found to be far from the quasi-exact result of Ref. \onlinecite{Cin10}, 
$\theta_{c}^{\textrm{mera}}\approx84.8^{\circ}$. On the other hand, 
the variational cluster approach gives a value of the critical angle 
$\theta_{c}^{\textrm{VCA}}\approx88^{\circ}$ much closer to 
$\theta_{c}^{\textrm{MERA}}$, however in both cases, i.e. in mean field 
and in the variational cluster approach, the result is potentially 
$U$-dependent, in agreement with earlier studies on the weak-coupling
limit.~\cite{0295-5075-97-2-27002}
As the variational cluster approach cannot directly detect nematic order 
by construction, we used here as a proxy the preferred spin direction, 
which is known to be different in the antiferromagnetic Ising and the 
nematic phases.~\cite{Cin10} However, at smaller $U\lesssim 8t$, where 
the space of plausible candidate phases is not known, the variational 
cluster approach results were inconclusive for large $\theta$. Bands at 
large $U$, where results are consistent, were interpreted with the help 
of mean field results.

We have obtained the spectral functions for the orbital compass model at 
different couplings $U$, as well as for the generalized compass model at 
strong coupling and different values of $\theta$. We compared these 
variational cluster-approach results with the mean field bands, where 
agreement at weak coupling, up to $U=2t$, is as expected good. 
(Only the metal-insulator transition occurred for smaller $U$ in case of 
the mean field.) For higher values of $U$, where the interacting spectra 
in the variational cluster approach become less coherent, agreement 
becomes worse. However, the bands obtained in mean field typically
still reproduce some features of the most coherent bands seen at the
top and the bottom of the spectra given by the variational cluster 
approach. 

The most striking feature of the orbital compass model, with respect to 
the corresponding itinerant model, is its nematic order, where chains 
with antiferromagnetic order stagger along one direction, say $a$, and 
are mutually decoupled along the other, $b$. 
The main topic of this paper is hole motion in such a phase. We have 
shown in Sec. \ref{sec:sym} that the same symmetries that decouple
orbital order between chains also render the kinetic Hamiltonian for
the hole independent of their relative orientation. As a consequence of 
the symmetry considerations, we can thus conclude that doping with a 
hole {\it does not lift} 
the degeneracy of the nematic ground state manifold. On a technical
side, this permits us to calculate one-particle spectra in one of the
ground states, e.g., the AF one, instead of having to average over many 
of them. This was confirmed by choosing different spin configurations
from the ground state manifold and getting the same result in the 
variational cluster approach. The disorder of the nematic ground state 
manifold does thus not affect hole motion and the variational 
cluster-approach spectra reveal rather coherent bands that also 
disperse along the $b$ direction, see Fig.~\ref{fig:spec_cmp_U2} in
Sec.~\ref{sec:spec}. This is in contrast to a spin-orbital model for
narrow-band manganites, where a nematic phase emerges spontaneously
without the Hamiltonian having similar symmetries, and where spectra
differ for different states.~\cite{Liang:2011fe}
For of a hole inserted into the $\alpha$ orbital, which
can hop along the AF ordered $a$ direction, comparison to mean field
reveals that three-site terms are crucial for the hole propagation, 
see Sec.~\ref{sec:spec}. This reflects the Ising-character of order 
along the $a$ axis, where there is moreover no inter-orbital hopping.
\cite{Dag08} 

In our study of the generalized compass model with two dimensional 
magnetic Ising order, we focused on the impact of interorbital hopping 
terms, parametrized by $\theta$, see Sec.~\ref{sec:hub}. At $\theta=0$,
interorbital hopping is absent and superexchange of the model is 
equivalent to the classical Ising model. The consequence for the hole's 
motion is clearly visible in the spectral function --- the bands are 
sharp but very flat. This is coherent with Ref. \onlinecite{Dag08}
saying that for classical Ising model the hopping of the hole is 
possible only via three-site terms. For $\theta>0$, where the 
strong-coupling limit no longer reduces to the Ising model, dispersion 
is strongly increased, even though the magnetic order is still 2D and 
Ising-like. Finite $\theta$ allows on one hand more quantum fluctuations 
into the ground state which enables hole propagation via spin-flips 
healing the defects produced by the hole, as in the Heisenberg model.
\cite{Mar91} On the other hand, inter-orbital hopping allows the hole 
to move even in an Ising-ordered background without quantum 
fluctuations, because it can hop without creating defects in the first 
place.~\cite{vdB00} This latter effect dominates in the generalized 
compass model, where order remains almost perfectly Ising-like.
\cite{Cin10}

The final conclusion on the mobility of a single hole in the above
models is that the hole can move coherently in the generic orbital 
compass model and its generalized version. In the Ising limit 
(at $\theta=0$ in the generalized compass model) as well as for the 
$\alpha$ orbital along the antiferromagnetic chains in $a$ direction 
of the orbital compass model, the dominant process is three-site 
hopping. Apart from this process, mobility in this latter case can be 
associated with the form of the hopping matrix $B_0$ 
in the direction of the $x$-bonds containing hopping between any pair 
of orbitals. This inter-orbital hopping allows the hole to avoid 
creating defects in the AF order by choosing the lowest-energy hopping
for each bond in the $b$ direction. However, this is {\it de facto}
more subtle for it would suggest coherent hopping to be only along the 
$b$ direction, while it is 2D in the variational cluster-approach 
spectra. In the generalized compass model, the analogous role is 
played by the $\sin^{2}(\theta/2)\sigma_{i}^{z}\sigma_{j}^{z}$ terms, 
see Eq. (\ref{eq:sigb}). This qualitatively explains, using essentially 
the same argument as for the generic orbital case, why the hole is 
confined in the generalized compass model when $\theta=0$, and becomes 
mobile when $\theta$ grows.

\acknowledgments
W.B. and A.M.O. kindly acknowledge support by the Polish National 
Science Center (NCN) under Project No. 2012/04/A/ST3/00331. 
W.B. acknowledges the kind hospitality of the Leibniz Institute 
for Solid State and Materials Research in Dresden.
M.D. thanks Deutsche Forschungsgemeinschaft
(grant DA 1235/1-1 under the Emmy-Noether program) for support.

\appendix

\section{General three-site hopping\label{sec:3site}}

The three-site hopping can be derived in the same way as the 
superexchange.
The only difference is that the fermions after creating a virtual
excitation with double occupancy do not come back to their initial
positions but move further on. In the Hamiltonians written below the
excitation was created at site $i$ and the fermion can either move
straight ahead either in the $a$ or $b$ directions, or can turn left 
or right after deexcitation. The general three-site hopping for the 
Hamiltonian ${\cal H}_{t-U}$ of Eq. (\ref{eq:H_tU}) reads,
\begin{eqnarray}
{\cal H}^{aa}_{t^{2}} & \!=\! & 
-\frac{t^{2}}{U}\sum_{i}\!\!\sum_{{\mu,\nu=\atop \alpha,\beta}}\!\!
\left\{ \tilde{c}_{i+a,\mu}^{\dagger}\!\left(\! 
A_{\nu\mu}A_{\nu\nu}\tilde{n}_{i,\bar{\nu}}\!+\! A_{\bar{\nu}\mu}
A_{\nu\bar{\nu}}\tilde{n}_{i,\nu}\!\right)\!\tilde{c}_{i-a,\nu}^{}
\right.\nonumber \\
& \!+\! & \left.\tilde{c}_{i+a,\mu}^{\dagger}\left[ 
A_{\bar{\nu}\mu}A_{\nu\nu}\!\left(\tilde{c}_{i,\nu}^{\dagger}
\tilde{c}_{i,\bar{\nu}}^{}\right)\!+\! A_{\nu\mu}
A_{\nu\bar{\nu}}\!\!\left(\!\tilde{c}_{i,\bar{\nu}}^{\dagger}
\tilde{c}_{i,\nu}\!\right)\!\right]\tilde{c}_{i-a,\nu}^{}\!\right\} 
\nonumber \\
 & \!+\! & {\rm H.c.},\label{eq:3s_bb_gen}
\end{eqnarray}
for the hopping along $a$ axis and analogical expression holds for
the $b$ axis. In case of turn at site $i$ the relevant expression
is,
\begin{eqnarray}
{\cal H}^{ab}_{t^{2}} & \!=\! & -\frac{t^{2}}{U}\!\sum_{i}\!\!
\sum_{{\mu,\nu=\atop \alpha,\beta}}\!\!\left\{ 
\tilde{c}_{i\pm b,\mu}^{\dagger}\!
\left(\! B_{\nu\mu}A_{\nu\nu}\tilde{n}_{i,\bar{\nu}}\!
+\! B_{\bar{\nu}\mu}A_{\nu\bar{\nu}}\tilde{n}_{i,\nu}\!\right)\!
\tilde{c}_{i\pm a,\nu}^{}\right.\nonumber \\
& \!+\! & \left.\tilde{c}_{i\pm b,\mu}^{\dagger}\!\left[
B_{\bar{\nu}\mu}A_{\nu\nu}\!\left(\tilde{c}_{i,\nu}^{\dagger}
\tilde{c}_{i,\bar{\nu}}^{}\right)\!+\! B_{\nu\mu}
A_{\nu\bar{\nu}}\!\left(\tilde{c}_{i,\bar{\nu}}^{\dagger}
\tilde{c}_{i,\nu}\!\right)\!\right]\tilde{c}_{i\pm a,\nu}^{}\!\right\} 
\nonumber \\
& \!+\! & {\rm H.c.},\label{eq:3s_ab_gen}
\end{eqnarray}
where $\bar{\alpha}(\bar{\beta})=\beta(\alpha)$. 

Now we can derive the three-site hopping Hamiltonians for the cases of 
OCM and GCM using hopping matrices of Eqs. (\ref{eq:hop_cmp}), 
(\ref{eq:Ahop_gcmp}) and (\ref{eq:Bhop_gcmp}). For OCM we get,
\begin{eqnarray}
{\cal H}^0_{t^{2}} & - & -\frac{t^{2}}{U}
\sum_{i}\tilde{c}_{i+a,\alpha}^{\dagger}
\tilde{n}_{i,\beta}\tilde{c}_{i-a,\alpha}^{}\nonumber \\
& - & \frac{t^{2}}{2U}\sum_{i}\!\sum_{\mu=\alpha,\beta}
\tilde{c}_{i\pm b,\mu}^{\dagger}\left(\tilde{n}_{i,\beta}
-\tilde{c}_{i,\alpha}^{\dagger}\tilde{c}_{i,\beta}\right)
\tilde{c}_{i\pm a,\alpha}^{}\nonumber \\
& - & \frac{t^{2}}{4U}\sum_{i}\!
\sum_{{\mu,\nu=\atop \alpha,\beta}}\!\tilde{c}_{i+b,\mu}^{\dagger}\!
\left(1-\sigma_{i}^{x}\right)\!\tilde{c}_{i-b,\nu}+{\rm H.c.},
\label{eq:cmp_3site}
\end{eqnarray}
with $\tilde{n}_{i,\beta}=n_{i,\beta}(1-n_{i,\alpha})$, and for the GCM,
\begin{eqnarray}
{\cal H}^{\theta}_{t^{2}} & = & 
-\frac{\sqrt{2}}{2}\frac{t^{2}}{U}\!\sum_{i} 
\sum_{{\mu,\nu=\atop \alpha,\beta}}\left\{ \frac{}{}\right.\nonumber \\
& \!\!\!\! & \tilde{c}_{i+a,\mu}^{\dagger} \left(\tilde{n}_{i}
-\sin\!\frac{\theta}{2}\sigma_{i}^{z}-\cos\!\frac{\theta}{2}
\sigma_{i}^{x}\right)A_{\mu\nu}^{\theta}\tilde{c}_{i-a,\nu}^{}\nonumber \\
& + & \tilde{c}_{i+b,\mu}^{\dagger}\left(\tilde{n}_{i} 
-\sin\frac{\theta}{2}\sigma_{i}^{z}
-\cos\frac{\theta}{2}\sigma_{i}^{x}\!\!\right) 
B_{\mu\nu}^{\theta}\tilde{c}_{i-b,\nu}^{}\nonumber \\
& + & \left.\tilde{c}_{i\pm b,\mu}^{\dagger}\!\!
\left(\sin\frac{\theta}{2}\tilde{n}_{i}-\sigma_{i}^{z}
-\cos\frac{\theta}{2}\sigma_{i}^{x}\sigma_{i}^{z}\right)
C_{\mu\nu}^{\theta}\tilde{c}_{i\pm a,\nu}^{}\right\} \nonumber \\
& + & {\rm H.c.},
\label{eq:gcmp_3site}
\end{eqnarray}
with $\tilde{n}_{i}=\tilde{n}_{i,\alpha}+\tilde{n}_{i,\beta}$ and
new hopping matrix $C^{\theta}$ similar to previous ones,
\begin{equation}
C^{\theta}=\frac{\sqrt{2}}{2}\left(\begin{array}{cc}
1+\sin\frac{\theta}{2} & \cos\frac{\theta}{2}\\
-\cos\frac{\theta}{2} & \sin\frac{\theta}{2}-1
\end{array}\right).
\end{equation}

\section{Change of basis for the $t$-$U$ model\label{sec:basis}}

We start with the $t$-$U$ Hamiltonian for the compass model of Eq.
(\ref{eq:H_cmp}),
\begin{eqnarray}
{\cal H}^0_{t-U} & = & t\sum_{i}\left\{ c_{i,\alpha}^{\dagger}
c_{i+a,\alpha}^{}+\frac{1}{2}\sum_{\mu,\nu=\alpha,\beta}
c_{i,\mu}^{\dagger}c_{i+b,\nu}^{}\right\} +{\rm H.c.}\nonumber \\
 & + & U\sum_{i}n_{i,\alpha}n_{i,\beta},
\end{eqnarray}
and we will transform it into the $t$-$U$ Hamiltonian of the GCM
Eq. (\ref{eq:H_gcmp}) at $\theta=\pi/2$. The key
transformation is rotation by $\pi/4$ in the fermionic space,
\begin{equation}
\left(\begin{array}{c}
c_{i,\alpha}\\
c_{i,\beta}
\end{array}\right)=\left(\begin{array}{cc}
\cos\frac{\pi}{8} & -\sin\frac{\pi}{8}\\
\sin\frac{\pi}{8} & \cos\frac{\pi}{8}
\end{array}\right)\left(\begin{array}{c}
b_{i,\alpha}\\
b_{i,\beta}
\end{array}\right).
\end{equation}
Following this one can express the hopping part in a new basis of
fermions $b_{\mu,i}^{\dagger}$ as,
\begin{eqnarray}
{\cal H}^0_{t} & \!=\! & \frac{t}{\sqrt{2}}\sum_{i}\!\sum_{\gamma=a,b}
\!\left\{\frac{\sqrt{2}\!+\!1}{2}b_{i,\alpha}^{\dagger}b_{i+\gamma,\alpha}^{}\!
+\!\frac{\sqrt{2}\!-\!1}{2}b_{i,\beta}^{\dagger}b_{i+\gamma,\beta}^{}\right.
\nonumber \\
& \!\mp\! & \left.\frac{1}{2}b_{i,\alpha}^{\dagger}b_{i+\gamma,\beta}^{}
\mp\frac{1}{2}b_{i,\beta}^{\dagger}b_{i+\gamma,\alpha}^{}\right\}.
\label{eq:hop_cmp_rot}
\end{eqnarray}
Surprisingly, the interaction part keeps its simple form after the
substitution, i.e.,
\begin{equation}
{\cal H}^0_{U}=U\sum_{i}\left(b_{i,\alpha}^{\dagger}
b_{i,\alpha}\right)\left(b_{i,\beta}^{\dagger}b_{i,\beta}\right).
\label{eq:int_cmp_rot}
\end{equation}
Comparing Eqs. (\ref{eq:hop_cmp_rot}) and (\ref{eq:int_cmp_rot})
with Eqs. (\ref{eq:H_tU}), (\ref{eq:Ahop_gcmp}) and (\ref{eq:Bhop_gcmp})
we see that the rotated $t$-$U$ compass model is equivalent to the
generalized $t$-$U$ compass model at $\theta=\pi/2$ if we only
renormalize the hopping amplitude $t$ by $1/\sqrt{2}$ in the compass
model.

\section{The Y antisymmetry\label{sec:y}}

The quantum compass model is know to anticommute with an operator
being a product of $\sigma_{i}^{y}$ on a chosen sublattice, i.e.,
\begin{equation}
Y=\prod_{i\in A}\sigma_{i}^{y}.
\end{equation}
Anticommutation means that,
\begin{equation}
Y{\cal H}_0^{J}Y=-{\cal H}_0^{J}.
\end{equation}
In the presence of a hole however, the $Y$ operator has to be modified
because $\sigma_{i}^{y}=0$ for a site with zero or double occupancy.
To cure this problem one can substitute $\sigma_{i}^{y}$ as follows
\begin{equation}
\sigma_{i}^{y}\to(1-n_{i})^{2}+\sigma_{i}^{y},
\end{equation}
with $n_{i}=n_{i,\alpha}+n_{i,\beta}$. Now at zero/double occupied
site, $\sigma_{i}^{y}=1$ and for other sites $\sigma_{i}^{y}$ remains
unchanged. The form of $Y$ for a single hole is,
\begin{equation}
Y_{1h}=\prod_{i\in A}\sigma_{i}^{y}+\sum_{p\in A}(1-n_{p})^{2}
\prod_{A\ni i\not=p}\sigma_{i}^{y}.
\end{equation}
Surprisingly, this does not change the anticommutation relation for
the pseudospin Hamiltonian, i.e.,
\begin{equation}
Y_{1h}{\cal H}^0_{J}Y_{1h}=-{\cal H}^0_{J},
\label{eq:H_cmp_bar}
\end{equation}
but the change in the kinetic part is less trivial,
\begin{eqnarray}
Y_{1h}{\cal H}^0_{t}Y_{1h} & = & \frac{t}{2}\,
\sum_{i\in A}\left(\tilde{c}_{i,\alpha}^{\dagger}
-\tilde{c}_{i,\beta}^{\dagger}\right)\!\sum_{\mu=\alpha,\beta}\!
\left(\tilde{c}_{i+b,\mu}\!+\!\tilde{c}_{i-b,\mu}\right)\nonumber \\
& - & t\,\sum_{i\in A}\tilde{c}_{i,\beta}^{\dagger}
\left(\tilde{c}_{i+a,\alpha} +\tilde{c}_{i-a,\alpha}\right)+{\rm H.c.}
\label{eq:H_cmp_t_bar}
\end{eqnarray}
These two results, Eqs. (\ref{eq:H_cmp_bar}) and (\ref{eq:H_cmp_t_bar})
show that on one hand the pseudospin interactions can be changed freely
from AF to FM, but on the other hand the kinetic part changes in such
a way that the physics of the moving hole remains unchanged. For 
instance, along $a$ the hopping transforms from pseudospin-conserving to 
pseudospin-flipping so the hole motion can frustrate the FM exchange in 
the $a$ direction.

\vfill
\eject

%


\end{document}